# Searching for Smurfs:
# Testing if Money Launderers Know Alert Thresholds

Rasmus Ingemann Tuffveson Jensen[a,b,*], Joras Ferwerda[c], Christian Remi Wewer[b]

[a]*Spar Nord Bank, Denmark,*
[b]*Department of Electrical and Computer Engineering, Aarhus University, Denmark,*
[c]*School of Economics, Utrecht University, the Netherlands,*

**Abstract**

**Objectives:** To combat money laundering, banks raise and review alerts on transactions that exceed confidential thresholds. However, the thresholds may be leaked to criminals, allowing them to break up large transactions into amounts under the thresholds. This paper introduces a data-driven approach to detect the phenomenon, popularly known as smurfing.
**Methods:** Our approach compares an observed transaction distribution to a counterfactual distribution estimated using a high-degree polynomial. We investigate the approach with simulation experiments and real transaction data from a systemically important Danish bank.
**Results:** Our simulation experiments suggest that the approach can detect smurfing when as little as 0.1-0.5% of all transactions are subject to smurfing. On the real transaction data, we find no evidence of smurfing and, thus, no evidence of leaked thresholds.
**Conclusions:** Our approach may be used to test if transaction thresholds have been leaked. This has practical implications for criminal justice and anti-money laundering (AML) systems. If criminals gain knowledge of AML alert thresholds, the effectiveness of the systems may be undermined. An implementation of our approach is available online, providing a free and easy-to-use tool for banks and financial supervisors. The null result obtained on our real data helps raise confidence in (though it cannot prove the effectiveness of) anti-money laundering efforts.

## 1. Introduction

The global anti-money laundering (AML) framework requires banks to report suspicious transactions to national authorities (FATF, 2021). To this end, some countries have publicly known reporting thresholds. For example, banks in the United States must, by law, report all currency transactions exceeding $10,000 (Glenn and Reed, 2023). In addition, banks are required to implement their own risk-based AML systems. These systems use confidential thresholds (set by each bank) to raise alerts. Bank officers review alerts and either (i) dismiss or (ii) report them to national authorities. The adversarial nature of money laundering, however, means that criminals change their *modus operandi* in response to AML efforts

---

[*]Corresponding author. Email: rasmus@tuffveson.com

(Welling, 1989). One straightforward strategy is known as smurfing or structuring (we use the terms interchangeably). The strategy splits up large transactions into multiple smaller ones (Truman and Reuter, 2004). In the United States, a money launderer may, for example, split up a large cash deposit into multiple smaller deposits below the publicly known $10,000 threshold. Banks are aware of this strategy and generally have confidential thresholds set below the national $10,000 threshold. However, with AML systems relying on confidential thresholds, banks and regulators may worry about these being leaked: indeed, there are multiple examples of bank employees helping criminals launder money (Klebnikov, 2022; Quintero, 2019; Roonemaa et al., 2022). If criminals discover the specific value of a confidential threshold, they may structure their transactions to remain below it, significantly reducing their probability of detection and, potentially, undermine the entire AML system.

In this paper, we present a data-driven approach to detect smurfing. The approach is inspired by the economic literature on bunching[1] and relies on the notion of a counterfactual distribution, i.e., an imagined distribution where money launderers do not adapt to alert thresholds. To test the approach, we employ data from Spar Nord, a systemically important Danish bank that (among other things) uses confidential alert thresholds. In this context, our approach can be seen as an attempt to detect if the thresholds have been leaked and are utilized by money launderers. We find no indication of smurfing and, thus, no indication that the bank's thresholds have been leaked. Our results are backed up by a Kolmogorov-Smirnov test, showing that transaction distributions did not change when the bank changed its threshold. Simulations, however, suggest that our approach can detect smurfing when as little as 0.1-0.5% of transactions are subject to smurfing.

Our paper is organized as follows. Section 2 discusses smurfing as a money laundering strategy. Section 3 discusses bunching in the economic literature. Section 4 describes our approach. Section 5 presents simulations to demonstrate the approach's utility. Section 6 applies the approach to real data. Finally, section 7 contains a conclusion and discussion.

## 2. The Challenge of Smurfing

The United States Congress passed the Currency and Foreign Transactions Reporting Act (also known as Title II of the Bank Secrecy Act) in 1970. The act requires banks to file reports on cash and coin transactions exceeding $10,000, aiming to provide authorities with reports of "(...) usefulness in criminal, tax, or regulatory investigations or proceedings" (Welling, 1989; United States Congress, 1982). However, by the mid-1980s, a problem with the act emerged: to avoid detection, criminals would structure large transactions into multiple smaller ones. Consequently, the United States Congress would go on to outlaw transaction structuring to avoiding the threshold in 1986, criminalizing money laundering (Van Duyne et al., 2018).

During the 1990s and early 2000s, AML regulation in both the United States and abroad shifted from a rule-based system to a risk-based system (Unger and van Waarden, 2009). Banks were now (and are still) expected to implement their own AML systems. Given that it is prohibitively expensive to review all transactions, the systems are designed to raise alerts on unusual or suspicious transactions: with the terms "unusual" and "suspicious" being intentionally vague. Alerts are further reviewed by bank officers before they, potentially, are reported to national authorities. However, there exists little research on AML systems, and authorities generally provide little feedback on their design. As a consequence, many systems still rely on simple thresholds like the original $10,000 rule. This leaves systems vulnerable to transaction structuring; their main defense being that their thresholds (supposedly) are confidential. Examples of bank insiders helping criminals, however, imply that confidential alert thresholds can give a false sense of security (Klebnikov, 2022; Quintero, 2019; Roonemaa et al., 2022). Furthermore, cash deposits still appear to be heavily applied in money laundering (Ardizzi et al., 2014).

---

[1]We thank Miroslav Palansky for this idea.

In this paper, we focus on smurfing schemes, i.e., money launderers splitting up large transactions into multiple smaller ones to decrease their chance of detection. A smurfing scheme can be as simple as a single money launderer making small cash deposits over a prolonged period of time. More advanced schemes may involve networks of low-level operatives known as smurfs or money mules, the purchase of goods and services, and complex firm ownership structures (Imanpour et al., 2019). To complicate matters further, smurfs may not be aware that they are participating in money laundering schemes (FBI, 2018; Europol, 2021).

The approach in our paper relies on two assumptions about smurfing:

1. it is not used to split up the very largest financial transactions, and
2. money launderers have incentives to make transactions close to (but always strictly below) alert thresholds.

The degree to which the first assumption holds will depend on the type of transactions being considered. For example, the probability that the very largest business transactions (e.g., the type made by large, publicly traded companies) would be money laundering seems low (at least in Denmark). However, money laundering can involve large transactions. Some studies and reports have, for example, indicated that illicit funds may be used to buy high-value assets like real estate and luxury goods (Financial Crimes Enforcement Network, 2006; Novaro et al., 2022; Teichmann, 2017). Thus, the validity of the assumption is open to discussion and will depend on the type of transactions being considered. Based on conversations with AML officers at the data-providing bank, we believe it is very unlikely that the very largest transactions of the types being considered in this paper are used in money laundering. In relation to the second assumption, we note that there is a discussion in the literature about how sophisticated and rational money launderers are (see Levi and Soudijn (2020); Truman and Reuter (2004); Ferwerda et al. (2020); Gelemerova (2011); Imanpour et al. (2019); Unger and den Hertog (2012); Nazzari (2024); McCarthy et al. (2015); Tiwari et al. (2023)). A perfectly rational money launderer would analyze and optimize their strategy. In particular, they would seek to minimize two types of costs. First, they would seek to minimize their expected punishment (i.e., their probability of being detected times the punishment they receive if detected). Second, they would seek to minimize smurfing costs (i.e., costs associated with splitting up large transactions and managing multiple smaller transactions). A large transaction is probably above any confidential threshold and therefore carries a high probability of detection. Multiple smaller transactions may be below the threshold but carry higher smurfing costs. A money launderer thus faces a trade-off. They may make a lot of very small transactions, but if these are far below the threshold, they incur unnecessary costs. Knowing a threshold value would be valuable for the money launderer; it would allow them to make transactions just below the threshold, lowering their probability of detection while minimizing their number of transactions and smurfing costs. This may undermine the entire AML system. In turn, a tool that can detect threshold leakage may be valuable for banks and financial supervisors.

## 3. Bunching in Economics

Smurfing may result in "bunching," a term long used by economists to describe how individuals tend to cluster their behavior just above or below thresholds where incentives change abruptly (e.g., around AML alert thresholds). Seminal studies on bunching consider taxation (Saez, 2010; Chetty et al., 2011; Kleven and Waseem, 2013), but bunching has also been studied in relation to insurance (Einav et al., 2017), fuel economy (Ito and Sallee, 2018; Sallee and Slemrod, 2012), and mortgage finance (Collier et al., 2021). The classic bunching setup considers a change in policy (e.g., the tax rate) at some continuous variable threshold (e.g., level of income), causing individuals to "manipulate" themselves below the threshold (e.g., report less income). A principal interest, then, is to find the fraction of manipulators. To this end, it is common to estimate a "manipulation-free" counterfactual distribution under two assumptions (Bachas et al., 2021):

1. manipulation is one-sided and bounded, and
2. the counterfactual distribution is "well-behaved", i.e., it can be estimated using non-manipulating individuals sufficiently far from the threshold.

The first assumption is reasonable in a smurfing context; see our discussion in section 2. The latter assumption, known as regularity, often translates to assuming that the counterfactual distribution is smooth across the threshold and can be estimated using a high-degree polynomial (Bachas et al., 2021). Whether this is reasonable in a smurfing context will depend on the particular type of transactions being considered. The (real) data that we consider in section 6 appear to follow log normal distributions quite closely, indicating that regularity is a reasonable assumption.

## 4. Detecting Smurfing with a Counterfactual Distribution

Our approach to detect smurfing relies on the notion of a counterfactual distribution, i.e., an imagined distribution where money launderers do not adapt to alert thresholds. The underlying idea is simple: we want to compare the (observed) empirical distribution with the (unobserved) counterfactual distribution. If the empirical distribution shows (i) an excess of transactions below a given threshold and (ii) a lack of transactions above the threshold, we consider it to be an indication of smurfing. In the subsections below, we describe the three steps of our approach:

1. create a transaction histogram in logarithmic space,
2. estimate a counterfactual distribution and calculate the relative amount of manipulated transactions, and
3. use bootstrapping to create confidence limits.

## 4.1. Creating a Transaction Histogram

Consider an alert threshold $T$ used to flag any transaction $t \geq T$. Motivated by log-normality of financial data, we assume $t, T > 1$ and log-transform and center any transaction $t$ around $T$, using the transformation

$$z(t) = \ln(t) - \ln(T). \tag{1}$$

We now want to construct a histogram where the alert threshold $z(T) = 0$ is a cutoff point (see figure 1 or 2 for an illustration). To make our approach robust to outliers, we only consider transactions with a size between the 0.001 and 0.999 percentiles. Let $\mathcal{T} = \{z(t_j)\}_{j=1}^{n}$ denote a set of transformed transactions after discarding transactions outside the percentiles. Note, in particular, that we let $n$ represent the number of observations. We use Doane's formula (Doane, 1976) to calculate an approximate number of bins $k$ for our histogram,

$$k = 1 + \log_2(n) + \log_2\left(1 + \frac{|g_1|}{\sigma_{g_1}}\right), \tag{2}$$

rounded to the nearest integer with $g_1 = \frac{1}{n}\sum_{j=1}^{n}\left(\frac{z(t_j)-\mu}{\sigma}\right)^3$, $\sigma_{g_1} = \sqrt{\frac{6(n-2)}{(n+1)(n+3)}}$, and $\mu$ and $\sigma$ denoting the mean and standard deviation of $\mathcal{T}$, respectively.[2] We then obtain a bin width

$$\omega = \frac{\max_{t_j \in \mathcal{T}} z(t_j) - \min_{t_j \in \mathcal{T}} z(t_j)}{k-1}, \tag{3}$$

and calculate

$$N_{\min} = \left\lfloor \frac{\min_{t_j \in \mathcal{T}} z(t_i)}{\omega} \right\rfloor \text{ and } N_{\max} = \left\lceil \frac{\max_{t_j \in \mathcal{T}} z(t_i)}{\omega} \right\rceil. \tag{4}$$

Finally, we construct bins $(b_i, b_{i+1}]$ by letting $b_i = i \times \omega$ for $i = N_{\min}, \ldots, -1, 0, 1, \ldots, N_{\max}$. We denote $(b_i, b_{i+1}]$ as bin $i$ and let $\mu_i = b_i + \frac{b_{i+1}-b_i}{2}$ denote the bin's middle point. To denote the fraction of transactions falling within each bin (relative to $n$), we use $n_i$.

## 4.2. Estimating the Amount of Manipulated Transactions

Let $U$ denote the maximum transaction size that a money launderer will seek to launder through smurfing and let $L$ denote the minimum transaction size involved in a smurfing scheme. We shall assume that these equal some bin cutoff points in our transaction histogram, i.e., that we have $z(L) = b_l < z(U) = b_u$ for some $l < 0 < u$. Based on the discussion in section 2, we have that:

1. a money launderer will split up a transaction of size $m \in [0, b_u)$ (in transformed space) into $d$ smaller transactions of size $s \in [b_l, 0)$, and
2. the empirical and counterfactual distributions are identical outside of $[b_l, b_u)$.

---

[2] Our implementation of Doane's formula is equivalent to an implementation in the Numpy library; see NumPy (2023).

To construct a counterfactual distribution, we now fit the fraction of observations $n_i$ as polynomial in the middle point $\mu_i$ where the empirical and counterfactual distributions are assumed to overlap, i.e., we fit

$$n_i = \beta_0 + \beta_1 \mu_i^1 + \cdots + \beta_p \mu_i^p \tag{5}$$

over all bins $i = N_{\min}, \ldots, l-1, u, \ldots, N_{\max}$, letting $\beta_0, \beta_1, \ldots, \beta_p \in \mathbb{R}$ denote an intercept and polynomial coefficients. For non-fitted bins, $i = l, \ldots, -1, 0, 1, \ldots, u-1$, we use estimates $\hat{n}_i$ based on (5) as a counterfactual distribution. We specifically calculate the proportion of manipulated transactions, accounting for the direction of manipulation, as

$$\zeta_{l,u} = \left( \sum_{i=l}^{-1} n_i - \hat{n}_i \right) + \left( \sum_{i=0}^{u-1} \hat{n}_i - n_i \right) \tag{6}$$

where $\hat{n}_i$ is our counterfactual estimate obtained from the fit of (5). The first term in (6) captures excess transaction below an alert threshold (indicating smurfing) while the latter term captures missing transactions above an alert threshold (also indicating smurfing). Thus, a substantial value $\zeta_{l,u} > 0$ implies the presence of smurfing.

### 4.3. Bootstrapping Confidence Limits

We use bootstrapping to calculate a lower confidence limit for $\zeta_{l,u}$ (see Ramachandran and Tsokos (2021) for a general reference). To be specific, we repeatedly:

1. draw $n$ samples with replacement from our data set $\mathcal{T}$, and
2. calculate $\zeta_{l,u}$ from the resampled data (keeping the bins from subsection 4.1 fixed but refitting the model from subsection 4.2).

The procedure is repeated 10,000 times, yielding 10,000 estimates of $\zeta_{l,u}$. The 5th percentile is then used as a lower 5% confidence limit. Our use of a one-sided confidence limit is motivated by the assumption that smurfing is one-sided, i.e., values of $\zeta_{l,u}$ significantly larger than 0 imply the presence of smurfing.

## 5. Simulation Experiments

We conduct simulation experiments to demonstrate the utility of our approach. We first simulate two baseline data sets (i.e., data sets without smurfing behavior). Both data sets are drawn from a normal distribution $\mathcal{N}(\mu, \sigma)$ using parameters estimated from our real transaction data (see section 6). Next, we artificially introduce smurfing into the two data sets. In all experiments, we fix $l = -1$ and let the polynomial degree $p$ of our model equal half the number of bins in any transaction histogram (rounded to the nearest integer; chosen for simplicity).

*5.1. Baseline Simulations*

To simulate baseline data, we employ parameters based on real transaction histograms (see section 6). We consider two different types of transactions, denoted as type A and B.[3] For type A, we draw 50,000 observations from a normal distribution with mean $\mu = -2.5$ and standard deviation $\sigma = 1.8$. For type B, we draw 250,000 observations from a normal distribution with mean $\mu = -2.1$ and standard deviation $\sigma = 2.1$. The means are based on fake but realistic alert thresholds (used to keep the real thresholds confidential). The standard deviations and numbers of observations directly reflect rounded statistics from the real data.

Figures 1, 2, 3, and 4 illustrate our polynomial model applied to the baseline simulated data.[4] Based on figure 1, it appears unreasonable to consider upper smurfing limits $u > 2$ for type A transactions; note how the fraction of observations in higher order bins is very small (less than 2% per bin). For type B transactions, we shall, however, consider upper smurfing limits up to and including $u = 3$. Table 1 contains estimates of $\zeta_{l,u}$ over both types of simulated baseline data. We find no evidence of smurfing. Using simulation parameters based on the real type A and B alert thresholds yield similar results (i.e., no model specifications indicate that smurfing is present).

|  | **Baseline (No Smurfing)** | | **Smurfing, $r = 0.1\%$** | | **Smurfing, $r = 0.5\%$** | |
|---|---|---|---|---|---|---|
|  | **Type A** | **Type B** | **Type A** | **Type B** | **Type A** | **Type B** |
|  | (0,0) | (0,0) | (49;106) | (249;554) | (249;534) | (1,247;2,807) |
| **u = 1** | -0.06 [-0.32] | -0.02 [-0.17] | 0.19 [-0.06] | 0.26 [0.12] | 1.20 [0.90] | 1.25 [1.08] |
| **u = 2** | -0.46 [-0.85] | -0.02 [-0.23] | -0.15 [-0.57] | 0.30 [0.10] | 1.05 [0.61] | 1.50 [1.29] |
| **u = 3** | - | -0.05 [-0.48] | - | 0.27 [-0.10] | - | 1.88 [1.47] |

**Table 1:** Estimates of $\zeta_{l,u}$ (in %) on simulated data with $l = -1$ and varying $u$ values. Lower 5%-confidence limit in squared bracket. Parentheses denote the number of money laundering transactions smurfed (i.e., removed) and induced (i.e., added) during simulation.

---

[3]We keep the transaction types confidential as to not disclose which transactions are subject to thresholds. However, one may, for example, think of cash deposits, wire transfers, or credit card purchases.

[4]For confidentiality, we only show plots of simulated data in this paper. Displaying plots of the real data, even if rescaled by a constant, might help adversaries infer the transaction types, A and B, being considered.

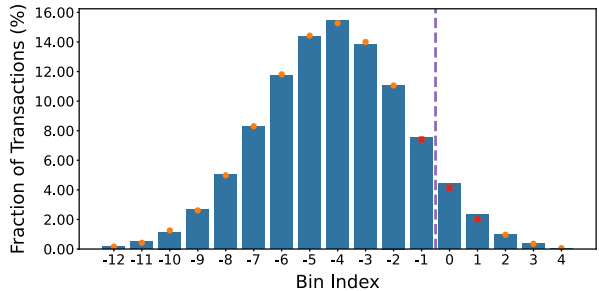

**Figure 1:** Type A baseline data using $l = -1$ and $u = 2$. Blue bars illustrate the fraction of transactions $n_i$ within each bin. Orange circles and red squares denote estimates $\hat{n}_i$ for fitted and non-fitted bins. Purple dashed line illustrates alert threshold.

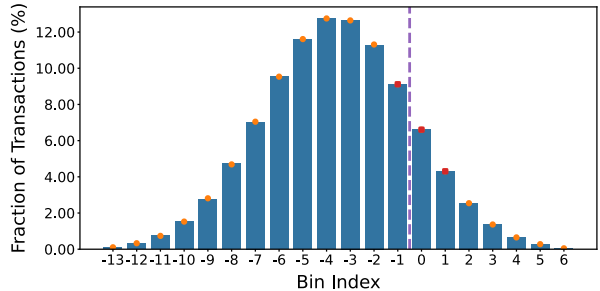

**Figure 2:** Type B baseline data using $l = -1$ and $u = 2$. Blue bars illustrate the fraction of transactions $n_i$ within each bin. Orange circles and red squares denote estimates $\hat{n}_i$ for fitted and non-fitted bins. Purple dashed line illustrates alert threshold.

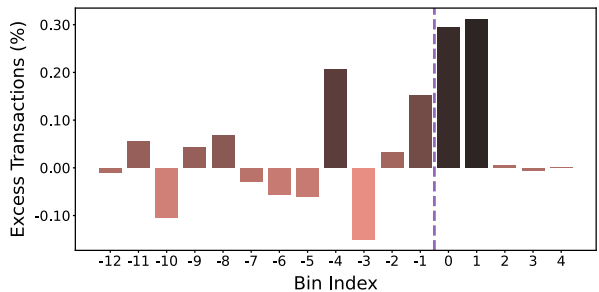

**Figure 3:** Excess transactions $(n_i - \hat{n}_i)$ on type A baseline data using $l = -1$ and $u = 2$. Darker colors reflect more excess transactions. Purple dashed line illustrates alert threshold.

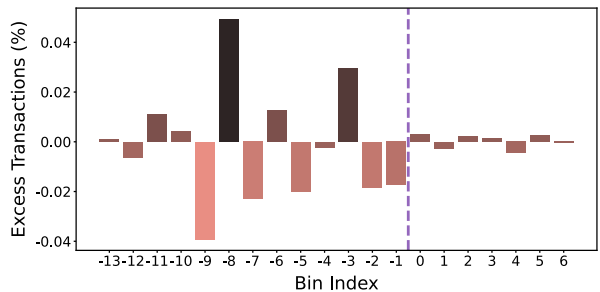

**Figure 4:** Excess transactions $(n_i - \hat{n}_i)$ on type B baseline data using $l = -1$ and $u = 2$. Darker colors reflect more excess transactions. Purple dashed line illustrates alert threshold.

### 5.2. Introducing Smurfing

To introduce smurfing into our simulated data, we fix $u = 2$ and $l = -1$. Let $n$ denote our number of transactions and $r \in [0, 1]$ some fraction of transactions to be smurfed. We randomly select $j = 1, \ldots, \lfloor r \times n \rfloor$ transactions of (transformed) size $m_j \in [0, b_u)$. For each selected transaction $j$ we:

1. draw a smurf transaction size $s_j$ from the uniform $U(b_l, 0)$ distribution,
2. calculate the number of size $s_j$ transactions permitted by $m_j$ implied by the logarithmic quotient rule,[5]

$$d_j = \lfloor \exp(m_j - s_j) \rfloor, \tag{7}$$

   and
3. add $d_j$ transactions of size $s_j$ to our data while removing transaction $j$.

By rounding down in (7), we imagine that a money launderer chooses not to launder the remainder between the amount to be laundered and the (untransformed) smurf transaction size. While we only select a fraction $r$ of all transactions to be smurfed, these will make up a larger fraction of the total transaction value (due to the fact that they are sampled in logarithmic space and in the $[0, b_u)$ interval).

---

[5]Let $x, y \in \mathbb{R}$ be two untransformed transaction sizes with $y < x$. The ratio between them (i.e., how many size $y$ transactions one can "split" $x$ into) can then be calculated as $x/y = \exp(\ln(x/y)) = \exp(\ln(x) - \ln(y))$.

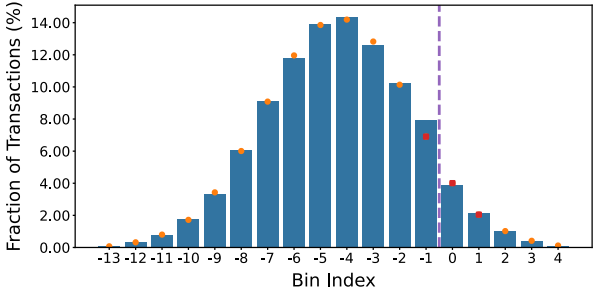

**Figure 5:** Type A simulated data with $r = 0.5\%$, $l = -1$, and $u = 2$. Blue bars illustrate the fraction of transactions $n_i$ within each bin. Orange circles and red squares denote estimates $\hat{n}_i$ for fitted and non-fitted bins. Purple dashed line illustrates alert threshold.

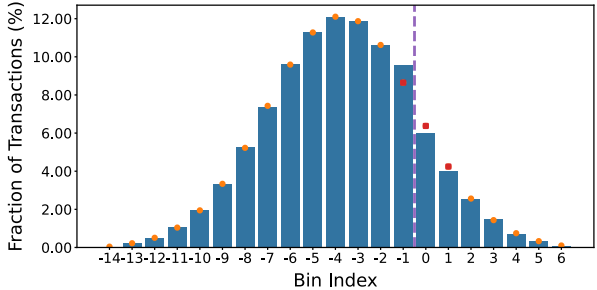

**Figure 6:** Type B simulated data with $r = 0.5\%$, $l = -1$, and $u = 2$. Blue bars illustrate the fraction of transactions $n_i$ within each bin. Orange circles and red squares denote estimates $\hat{n}_i$ for fitted and non-fitted bins. Purple dashed line illustrates alert threshold.

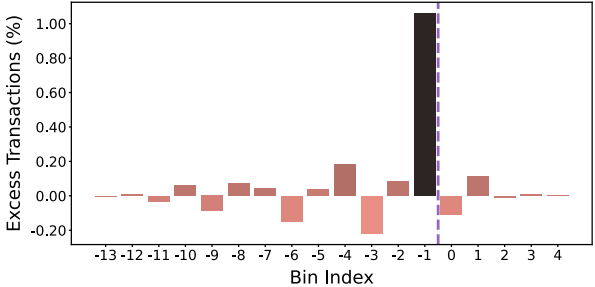

**Figure 7:** Excess transactions $(n_i - \hat{n}_i)$ on type A simulated data with $r = 0.5\%$, $l = -1$, and $u = 2$. Darker colors reflect more excess transactions. Purple dashed line illustrates alert threshold.

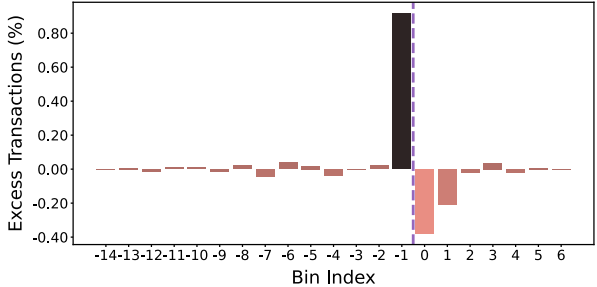

**Figure 8:** Excess transactions $(n_i - \hat{n}_i)$ on type B simulated data with $r = 0.5\%$, $l = -1$, and $u = 2$. Darker colors reflect more excess transactions. Purple dashed line illustrates alert threshold.

We first consider the case where $r$ corresponds to 0.5% of all transactions $n$ and apply the same test scheme as with the baseline data. Figures 5, 6, 7, and 8 display illustrations while table 1 contains estimates of $\zeta_{l,u}$. On both the type A and type B data, all model specifications detect the introduced smurfing behavior. We note that using simulation parameters based on the real type A and B alert thresholds yield similar results (i.e., all model specifications detect the introduced smurfing behavior).

We next consider the case where $r$ corresponds to 0.1% of all transactions $n$. Table 1 contains estimates of $\zeta_{l,u}$. On the type A data, no model specification detects smurfing. However, on the type B data, two out of three model specifications detect smurfing. Using simulation parameters based on the real type A and B alert thresholds yield similar results (i.e., the same model specifications, respectively, miss and detect the introduced smurfing behavior).

## 6. An Application to Real Data

We test our approach on real data from Spar Nord, a systemically important Danish bank with approximately 440,000 clients. We consider two types of transactions, denoted as type A and B, subject to different alert thresholds and associated with different client subgroups.[6] Unless explicitly stated, all transactions are collected over the year 2019. Transactions of type A are relatively sparse with around $50,000$ observations. Transactions of type B are more numerous with around $250,000$ observations. As in our simulation experiments, we fix $l = -1$ and let the degree $p$ of our polynomial model equal half the number of bins in each histogram (rounded to the nearest integer).[7]

*6.1. Results*

Table 2 displays estimates of $\zeta_{l,u}$ for our type A and type B transactions. The table contains estimates using both the real and fake alert thresholds (the latter being identical to the fake thresholds used in section 5). Note that we, as in section 5, consider $l = -1$ for both type A and B transactions, $u \leq 2$ for type A transactions, and $u \leq 3$ for type B transactions. Our results do not support the presence of smurfing.

|  | Type A (Real) | Type A (Fake) | Type B (Real) | Type B (Fake) |
|---|---|---|---|---|
| **u = 1** | -0.97 [-1.25] | -0.53 [-0.76] | 0.12 [-0.01] | -0.15 [-0.28] |
| **u = 2** | -3.22 [-3.61] | -1.40 [-1.72] | -0.27 [-0.45] | 0.05 [-0.22] |
| **u = 3** | - | - | -0.46 [-0.83] | 0.29 [-0.23] |

**Table 2:** Estimates of $\zeta_{l,u}$ (in %) on type A and B transactions with both real and fake alert thresholds, $l = -1$, and varying $u$ values. Lower 5%-confidence limit in squared bracket.

*6.2. A Kolmogorov-Smirnov Test*

The alert threshold associated with type A transactions was decommissioned sometime in 2020-2021. This allows us to use a Kolmogorov-Smirnov test as a complementary test for smurfing. The test quantifies the maximum distance between the empirical distribution functions of two samples. Under the null hypothesis, the two are identically distributed. Our idea, then, is to compare the distribution of transactions before and after the alert threshold was decommissioned. To this end, we first collect a sample of all type A transactions made one month before and up until the alert threshold was decommissioned. Next, we collect a sample of all type A transactions made over the same period one year later. We obtain a $p$-value equal to 0.2579 and, thus, find no indication of smurfing. A similar test using 2 months of data yields a $p$-value of 0.2611. Going back more than 2 months means collecting data from a period where Denmark was in Covid-19 lockdown. Doing so, we consistently obtain $p$-values below 0.05, indicating that samples do not come from identical distributions. We stress that we do not believe this to be a result of smurfing; rather, we have strong reason to believe that the Covid-19 lockdown influenced the distribution of type A transactions. Indeed, the lockdown influenced the distribution of many types of financial transactions, including, for example, retail consumer spending and international trade flows due to reduced in-person activity and disrupted spending and payment patterns (Berthou and Stumpner, 2024; Chen et al., 2024; Rose et al., 2023).

---

[6] Due to confidentiality, we cannot disclose the exact types of considered transactions. However, the reader may, for example, think of cash deposits, wire transfers, or credit card transactions.

[7] We stress that Spar Nord applies additional logic such that not all transactions over the thresholds are flagged; as the reader might imagine, this would otherwise yield a prohibitively high number of alerts. Furthermore, the bank employs additional rules, models, and scenarios that may flag transactions below thresholds.

## 7. Discussion and Conclusion

We present an approach to detect smurfing, a particular type of money laundering where money launderers split large transactions into multiple smaller transactions below an alert threshold. Our approach employs the notion of a counterfactual distribution and relies on two assumptions about smurfing: (i) it is not a viable option to launder very large amounts of money and (ii) money launderers have incentives to make transactions just below alert thresholds. The assumptions remain untested in our work. Simulation experiments suggest that our approach can detect smurfing when as little as 0.1-0.5% of transactions are subject to smurfing. An application to real data finds no evidence of smurfing and, thus, no evidence of threshold leakage. The results are backed up by a Kolmogorov-Smirnov test.

Not all real transaction data will follow log-normal distributions as closely as the data considered in our application. This may impact the performance of our approach. In addition, our simulation results are, naturally, specific to the employed simulation parameters (manifesting from underlying transaction types and employed alert thresholds). Note, in particular, that our simulations induce a relatively small number of new transactions per smurfed money laundering transaction. Thus, we recommend applications of our approach be accompanied by simulations using case-specific parameters.

Our approach and findings have practical implications for criminal justice and AML systems. If criminals gain knowledge of AML alert thresholds, the effectiveness of AML systems may be undermined. While we find no evidence of threshold leakage in our real data, testing for leakages remains important in and of itself to ensure the effectiveness of AML systems. Our approach may be used by banks and financial supervisors alike to gain confidence in AML efforts.

**CRediT Author Statement**

**Rasmus Ingemann Tuffveson Jensen**: Conceptualization, Methodology, Formal Analysis, Writing - Original Draft. **Joras Ferwerda**: Conceptualization, Writing - Review & Editing, Supervision. **Christian Remi Wewer**: Software, Writing - Review & Editing.

**Data Availability**

Data is not available due to the confidential nature of financial transactions.

**Supplementary Material**

An implementation of our approach (including our simulation experiments) is available on GitHub: `https://github.com/TuffvesonJensen/Searching4Smurfs`.

## Declaration of Interest

The authors declare no competing interests.

## Acknowledgements

We are grateful to an anonymous AML data scientist at a major Danish bank (not Spar Nord) for feedback on our manuscript and code.